\documentclass[aps,twocolumn,showpacs]{revtex4}
\usepackage{graphicx}
\usepackage{amsmath}
\usepackage{amssymb}
\usepackage{bm}

\begin{document}

\title{Duality in the Kondo model and perturbative approach to strong coupling theory}

\author{Tomosuke Aono}
\affiliation{%
Department of Physics,
Ben-Gurion University of the Negev,
Beer-Sheva 84105, Israel}
\date{\today}
%

\begin{abstract}
We investigate the duality in the Kondo model.
Starting from the s-d model with
the coupling constant $J$,
the strong coupling model with the constant of $1/J$ is identified.
The model shows the unitary limit of the conductance, $G=2 e^{2}/h$ at zero temperature.
The perturbation theory of the model gives qualitative agreement with the results of numerical renormalization group and Bethe ansatz near $T=T_{K}$.
\end{abstract} 

\maketitle


Ever since Kondo elucidated the origin of logarithmic divergence of
the resistivity in dilute alloy systems~\cite{Kondo64},
various aspects of the Kondo effect have been investigated
\cite{Hewson92},
including recent experimental realizations 
in quantum dot systems~\cite{Goldhaber98,Cronenwett99,Simmel99,Wilfred00,Sasaki00,Ji00,Jeong01,Craig04}.
When the temperature or a characteristic energy scale is
larger than the Kondo temperature $T_{K}$,
the system is in the weak coupling regime while
it is smaller, the system is in the strong coupling regime.
In these regimes, the Kondo model exhibits duality.
It was pointed out from the exact solution~\cite{Fateev81}.
In Refs.~\cite{Fendley98,Lesage99},
it was revealed for several quantum impurity problems,
including $S=1/2$ Kondo problem;
In particular,
the resistivity has calculated beyond the Fermi liquid regime~\cite{Lesage99}, from the point of view of the duality.
The duality also plays an important role
in quantum dissipative systems
\cite{Schmid83,Fisher85,Weiss99},
which are closely related with quantum impurity problems.
It is possible to approach
perturbatively  for a strong coupling regime
using a similar way in a corresponding weak coupling regime.

We are presenting yet another duality for
$S=1/2$ Kondo model with the coupling constant $J$.
The motivation of this work comes from the following two points.
The first one is a recent active interest on the Kondo screening cloud~\cite{Nozieres75,Sorensen95,Barzykin96,Ujsaghy00}.
At low temperatures, conduction electrons and the impurity form
the Kondo singlet state, accompanying  a screening cloud.
The nature of the cloud is still under debate,
and accordingly many works have been investigated
to detect the cloud in nano structures~\cite{Zvyagin97,Ferrari99,Affleck01,Hu01,Eckle01,Simon02,Cornaglia03,Kang00,Sorensen05}.
The second one  is the notion of the strong coupling theory itself.
The Fermi liquid theories of the Kondo effect~\cite{Nozieres74, Yamada75}
have been developed and they successfully explain low temperature behavior of
the Kondo effect~\cite{Hewson92}.
Although it is stated that
strong coupling  theories describe
the low temperature regime,
simple questions have  not yet answered clearly;
What is the action (Hamiltonian)  with the coupling constant of
$1/J$ if it exists?,
and 
how does the perturbation with respect to $1/J$ work in the theory?

We will explain that the concept of the screening cloud naturally leads to
a duality in the Kondo problem.
We will identify the model which has a very large cloud 
starting from the conventional weak coupling s-d model
and show it  has the coupling constant of $1/J$.
Electron transport of this model will be investigated,
which shows the unitary limit of the conductance at zero temperature.
A perturbative  calculation with respect to $1/J$ with
the poorman's scaling is performed,
giving good qualitative agreement with the results of numerical renormalization group (NRG)~\cite{Costi94} and Bethe ansatz~\cite{Konik02}
around $T =T_{K}$.

{\it Weak coupling model.---}
We start from the  s-d model.
The action $S = S_0 + S_1$ is 
\begin{eqnarray}
 S_0= \int d\omega\sum_{k,\sigma}\bar{\psi}_{k\sigma}(\omega)
\left(\omega-\epsilon_{k}\right)\psi_{k\sigma}(\omega),
\end{eqnarray}
\begin{eqnarray}\label{eq:s-d_interaction}
 S_1 = J \int dt\left[
\begin{array}{cc}
\bar{\psi}_{\uparrow}(t) & \bar{\psi}_{\downarrow}(t)\end{array}\right]
M
\left[\begin{array}{c}
\psi_{\uparrow}(t)\\
\psi_{\downarrow}(t)\end{array}\right],
\end{eqnarray}
with
\begin{eqnarray}
M =\left(\begin{array}{cc}
\frac{1}{2}S_{z} & \frac{1}{2}S_{-}\\
\frac{1}{2}S_{+} & -\frac{1}{2}S_{z}\end{array}\right),
\end{eqnarray}
where 
$\psi_{k \sigma}$ is the conduction electron field with the momentum $k$ and spin $\sigma$.
We have introduced the surface field operator
$ \psi_{\sigma} = \sum_k \psi_{k \sigma}$,
which is the conduction electron field at the impurity site ($x=0$).
For simplicity, we have disregarded the potential scattering term.

The two conduction electron fields, $\psi_{\sigma}$ and $\psi_{k\sigma}$ 
can be treated independently
if we introduce a Lagrange multiplier $\lambda_{\sigma}$ and
add a constraint term to the action:
\begin{eqnarray}
\int dt
\sum_{\sigma} \bar{\lambda}_{\sigma}(t) 
\left( \psi_{\sigma}(t) - \sum_{k} \psi_{k \sigma} \right)
+
{\rm H.c.}
\end{eqnarray}
The action can be written only by
the surface field $\psi_{\sigma}$ after
integrating over $\psi_{k \sigma}$ and $\lambda_{\sigma}$:
\begin{eqnarray}\label{eq:weak-coupling-action}
  S = \sum_{\sigma}\int d\omega \bar{\psi}_{\sigma}(\omega) G^{-1}(\omega) \psi_{\sigma}(\omega) 
  + S_1,
\end{eqnarray}
where $G(\omega)$ is the Green function of conduction electrons.
At zero temperature, $G(\omega) = i/2\; (2 \pi \nu)\; {\rm sgn} (\omega)$ with
the density of states $\nu$ in the lead and the sign function ${\rm sgn}(x)$.
For finite temperature,
it is convenient to introduce
the Keldysh Green function (matrix)
\begin{eqnarray}
  \label{eq:Keldysh_Green_Functions}
  G(\omega)=\left(\begin{array}{cc}
 2 \pi i \nu[\frac{-1}{2}+f(\omega)] &  2 \pi i \nu f(\omega)\\
 2 \pi i \nu[f(\omega)-1] &  2 \pi i \nu [\frac{-1}{2}+f(\omega)]\end{array}\right)
\end{eqnarray}
where
the matrix is defined on the Keldysh space
[(1,1) component of the matrix gives the time-ordered Green function, for example.]

{\it Strong coupling model.---}
Now we show the duality of the Kondo model.
Instead of integrating over $\lambda_{\sigma}$, we integrate over $\psi_{\sigma}$ and $\psi_{k\sigma}$,
and then the action is represented by $\lambda_{\sigma}$ in the form:
\begin{eqnarray}\label{eq:strong-field-action}
&& S = \int d\omega \bar{\lambda}_{\sigma}(\omega) g^{-1}(\omega)
\lambda_{\sigma}(\omega) \nonumber\\
&&  +
  \frac{16}{3J} \int dt
\begin{pmatrix}
\bar{\lambda}_{\uparrow} & \bar{\lambda}_{\downarrow}
\end{pmatrix}
M^{-1}
\begin{pmatrix}
\lambda_{\uparrow}\\
\lambda_{\downarrow}
\end{pmatrix}
\end{eqnarray}
with
\begin{eqnarray}
M^{-1}=
\left(
\begin{array}{cc}
\frac{1}{2}S_{z} + \frac{1}{2} & \frac{1}{2}S_{-}\\
\frac{1}{2}S_{+} & -\frac{1}{2}S_{z}+\frac{1}{2}
\end{array}
\right),
\end{eqnarray}
where we have used the identity of $(\vec{\sigma} \cdot \vec{S})
(\vec{\sigma}
\cdot
\vec{S})  = S^{2} + i \vec{\sigma} \cdot \vec{S} \times \vec{S}$, and 
$\vec{S} \times \vec{S} = i \vec{S}$.
The Green function $g(\omega)$ is the inverse of $G(\omega)$.
For example, $g(\omega) =-2i/(2 \pi \nu)\; {\rm sgn} (\omega)$ for zero temperature, and
the Keldysh Green function is
\begin{eqnarray}
g(\omega)=\left(\begin{array}{cc}
\frac{-4i}{2 \pi \nu}[\frac{-1}{2}+f(\omega)] & \frac{4i}{2 \pi \nu}f(\omega)\\
\frac{4i}{2 \pi \nu}[f(\omega)-1] & \frac{-4i}{2 \pi
\nu}[\frac{-1}{2}+f(\omega)]\end{array}\right).
\end{eqnarray}
Equation (\ref{eq:strong-field-action}) looks quite similar to
the weak coupling model;
It includes the s-d interaction as
in Eq.~(\ref{eq:s-d_interaction}) with the coupling constant
proportional to $1/J$ instead of $J$.
It also contain an additional "potential scattering" term which
comes from the commutation relation among spin operators.
We call this model as the strong coupling Kondo model, and
define $j = 16/(3 J)$.

Now we discuss the procedure has been made.
The role of the Lagrange multiplier is to enforce the surface field $\psi_{\sigma}$ to locate at the impurity site;
The interaction with the impurity is allowed only at the origin.
As the temperature decreases,
conduction electrons start to screen the impurity,
accompanying  the screening cloud,
which is the sign of the Kondo effect.
The size of the cloud is $\sim v_{F}/ T_{K}$ with
the Fermi velocity $v_{F}$ of conduction electrons.
It is estimated~\cite{Affleck01} to be the order of $1 \mu m$ in a recent experiment~\cite{Wilfred00};
It is much larger than the size of the quantum dot.
The large cloud means that
conduction electrons away from the origin can interact with the impurity strongly.
Thus the surface field,
which interacts with the localized spin,
is no longer unnecessary to locate at the origin.
As a result, $\psi_{\sigma} - \sum_{k} \psi_{k\sigma}$ fluctuates
when the cloud is formed,
and accordingly should be integrated out from the action.
This means that the conjugate field, $\lambda_{\sigma}$ now describes the low energy physics instead.
In this sense, the strong coupling model (\ref{eq:strong-field-action}) will describe the low temperature regime of the Kondo effect.
In the following, this point is examined.

{\it Poorman's scaling.---}
Next we discuss the scaling of $j$ of the strong coupling model
using the poorman's scaling as in the weak coupling theory~\cite{Anderson70}.
To begin with this,
we includes the "potential scattering" term as the zeroth order action.
Then the time-ordered Green function is
\begin{eqnarray}
g(\omega) = \frac{-4i}{2 \pi \nu}[\frac{-1}{2}+f(\omega)] \frac{1}{1+(j/2\pi\nu)^2}
+ \frac{(j/2\pi\nu)}{1+(j/2\pi\nu)^2}.
\end{eqnarray}
We perform the second order perturbation expansion with respect to $j$
and then
the RG equation reads
\begin{eqnarray}\label{eg:RG-exact}
  \frac{d j}{d \log D} =  (+4/(4 \pi^2 \nu))   \frac{j^2}{1+(j/2\pi\nu)^2}.
\end{eqnarray}
with an energy cut-off $D$.
The second term in the time-ordered Green function does not
induce the logarithmic contribution.

There is an important difference in this RG;
the sign of the right hand side is the opposite to the one in the 
weak coupling model.
It comes from the difference in 
the sign of the time-ordered Green function.
This result means that $j$ becomes smaller
as $D \rightarrow0$;
The s-d interaction is less prominent at low temperatures,
and eventually vanishes at $T=0$.
In terms of the weak coupling model, $j=0$ corresponds to
$J \rightarrow \infty$,
a strong coupling fixed point.
Note that
the scaling flow is formally the same as in the ferromagnetic Kondo model.
The scaling flow allows to disregard $j$
in the denominator in Eq.~(\ref{eg:RG-exact})
which comes from the potential scattering term.
Thus we disregard this term in the following.
Then the solution is
\begin{eqnarray}\label{eq:j-strong}
  j(D) =  \frac{j_0}{1+ 1/\nu \pi^2 j_0 \log{d_0/D}},
\end{eqnarray}
where $j_0 = j(d_0)$ with the initial cut-off $d_0$.

To make a connection between the weak and strong coupling theories,
$j_0$ and $d_{0}$ should be represented by $J_{0}$ and $D_{0}$.
The first  condition is 
$j(d_{0})/\nu = 16/3 ( 1/\nu J(d_{0}))$ with
$\nu J(d_{0}) = 1/ \log(d_{0}/T_{K})$.
($d_{0}$ should be larger than $T_{K}$.)
This results in
\begin{eqnarray}\label{eq:j-scaling}
j_{0}(D)/\nu = \frac{1}{\frac{3}{16 \log(d_{0}/T_{K})} + \frac{1}{\pi^{2}} \log(d_{0}/D)}.
\end{eqnarray}
The value of $d_{0}$ is still arbitrary, and the second condition is required.
Equation (\ref{eq:j-scaling}) diverges at 
$D=d_{1}$, which is a function of $d_{0}$.
We impose the condition $\partial d_{1}/\partial d_{0}=0$
because the divergence should not depend on
the artificial parameter $d_{0}$.
This gives $d_{0} = \exp (\sqrt{3/16} \pi) T_{K} \sim 1.36 T_{K}$.
The  two conditions eventually yield
\begin{eqnarray}\label{eq:coupling-constant-strong}
 j(D)/\nu =  \frac{\pi^{2}}{\log(T_{K}^{*}/D)}} \;\;{(D <  d_{0}).
\end{eqnarray}
with $T^{*}_{K}=  \exp (2 \pi \sqrt{3/16}) T_{K}
\sim 15.1 T_{K}$.
Since $T_{K}^{*}$ is larger than $T_{K}$,
$j$ is a smooth function of $D$ near $T=T_{K}$
while  $J$ diverges at $T_{K}$.
This provides a contrast between two perturbation theories as will be shown below.

{\it The expression of the current.---}
We shift our discussion to electron transport, and
consider the model with the impurity attached to 
two leads $\alpha=L,R$ with a chemical potential difference $\mu$.
The Hamiltonian is $H_0+ H_1$ with
$H_0 = \sum_{\alpha,k,\sigma}
  \bar{\psi}_{\alpha k\sigma} (\omega \pm \mu/2 -\epsilon_k)
\psi_{\alpha k\sigma}$, [the sign of $+$($-$) is for L(R)], and
$H_1 = J \sum_{\alpha, \alpha'} \bar{\psi}_{\alpha}  \vec{S} \cdot \vec{\sigma}
\psi_{\alpha'}$. 
We have assumed fully symmetric constants between $\alpha = L,R$
for simplicity.
The model is simplified, when we apply 
a unitary transformation between the conduction electron fields,
$\varphi_{(k)\sigma}=1/\sqrt{2}(\psi_{L(k)\sigma}+\psi_{R(k)\sigma})$ and
$\psi_{(k)\sigma}=1/\sqrt{2}(\psi_{L(k)\sigma}-\psi_{R(k)\sigma})$;
The $\psi_{(k)\sigma}$ field is decoupled from
the impurity, and it simply represents free electron gas.
A similar transformation is used
in the Anderson model~\cite{Glazman88}.
We discuss the conductance  through the impurity.
The current operator $I(t)$
$\frac{i e}{\hbar} [ \bar{\psi}_L(t) \vec{S} \cdot \vec{\sigma} \psi_R(t) - \bar{\psi}_R(t) \vec{S} \cdot \vec{\sigma} \psi_L(t) ]$
is also rewritten by
\begin{eqnarray}
I = \frac{ie}{\hbar} J [\bar{\varphi} \vec{S} \cdot \vec{\sigma} \psi - \bar{\psi} \vec{S} \cdot
\vec{\sigma} \varphi ].
\end{eqnarray}
This term is added to the action in the form of $\int dt \eta(t) I(t)$.

We repeat a similar procedure as above,
rewriting the action using
the Lagrange multiplier $\lambda_{\alpha}$.
Applying the unitary transformation for $\lambda_{\alpha}$,
$\lambda_{\varphi} = 1/\sqrt{2} (\lambda_L + \lambda_R)$ and
$\lambda_{\psi}=  1/\sqrt{2} (\lambda_L - \lambda_R)$, 
the two lead s-d model
including  the current generating term is given by
\begin{eqnarray}
S&=& \int d\omega 
\begin{pmatrix}
\bar{\lambda}_{\varphi} & \bar{\lambda}_{\psi}
\end{pmatrix}
\begin{pmatrix}
A & B\\
B & C
\end{pmatrix}
\begin{pmatrix}
\lambda_{\varphi} \\
\lambda_{\psi}
\end{pmatrix} \nonumber\\
&&
+ \int d\omega\; [\bar{\psi} \lambda_{\psi} + \bar{\lambda}_{\psi} \psi]
+ \int dt\; [\bar{\varphi} \lambda_{\varphi} + \bar{\lambda}_{\varphi} \varphi]
\nonumber \\ && + (2J) \int dt\; \bar{\varphi} M \varphi 
 +  e J \int dt\; \eta(t) [ \bar{\varphi} M \psi - \bar{\psi} M
\varphi]\nonumber\\
\end{eqnarray}
with
$A = C=
\begin{pmatrix}
\frac{-i}{2} (2 \pi \nu) & \frac{1}{2}(f_K+g_K)\\
0 & \frac{i }{2}(2 \pi \nu),
\end{pmatrix},
$
$
B=
\begin{pmatrix}
 0 & \frac{-1}{2}(f_{K}-g_{K})\\
 0 & 0
\end{pmatrix},
$
and $f_{K}=-i (2 \pi \nu) \tanh(\omega-\mu/2)/2$
and $g_{K}=-i (2 \pi \nu) \tanh(\omega+\mu/2)/2$.
In these matrices, we have represented them in the Keldysh rotation representation.

The low energy action should be represented by 
$\lambda_{\varphi}$ and $\psi$ because 
the $\varphi$ field participates in the screening while
the $\psi$ field does not.
Accordingly integrating over $\varphi$ and $\lambda_{\psi}$,
we obtain
\begin{eqnarray}
&&S= \int d\omega 
\begin{pmatrix}
\bar{\lambda}_{\varphi} & \bar{\psi}
\end{pmatrix}
\begin{pmatrix}
a & b\\
-b& c
\end{pmatrix}
\begin{pmatrix}
\lambda_{\varphi} \nonumber \\
\psi
\end{pmatrix}\\
&&+ \frac{16}{3 (2J)} \int dt\; \bar{\lambda}_{\varphi} M^{-1} \lambda_{\varphi}
+ \frac{e}{2} \int dt\;  \eta(t) [\bar{\lambda}_{\varphi} \psi - \bar{\psi} \lambda_{\varphi}]\nonumber \\
\end{eqnarray}
with
$a=c^{-1}=A$, 
$b = \begin{pmatrix}
0& \frac{i(f_K-g_K)}{2 \pi \nu}\\
0 & 0
\end{pmatrix}$.
The important result of this action is that
the current generating term represented
by $\lambda_{\varphi}$ and $\psi$
is independent on $J$ and has no spin dependence
although initially it depends on $J$ and the spin.
The lowest order of the conductance 
together with the result of the RG gives
the unitary limit of $G=2 e^{2}/h$.

{\it Strong coupling perturbation theory.---}
We apply the perturbation theory with respect to $1/J$.
The leading and next to leading order contributions give
\begin{eqnarray}\label{eq:conductance-perturbation}
 G =\frac{2 e^2}{h}  \left[ 1- \frac{3 j^2}{8(2 \pi\nu)^2} \right]
\end{eqnarray}
while $G=\frac{2 e^2}{h} (2 \pi \nu J)^2\frac{3}{16}$ for the weak coupling model~\cite{Kaminski99}.
We have disregarded the potential scattering term.
To include the Kondo effect,
the coupling constants should be replaced by
the renormalized ones; $J \rightarrow J(D)$ and $j \rightarrow j(D)$.

\begin{figure}
  \includegraphics[width=\columnwidth]{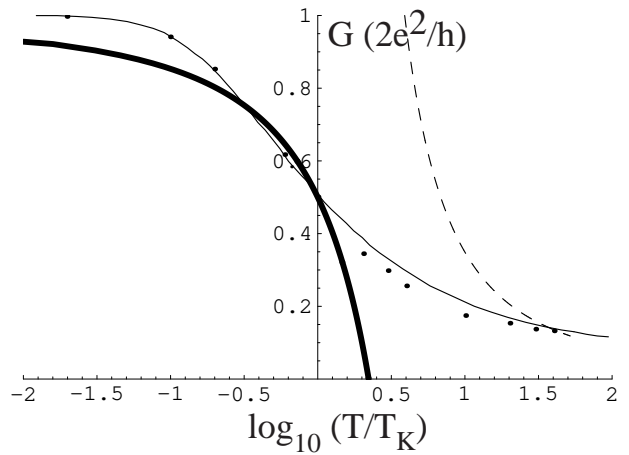}
  \caption{%
The conductance as a function of $\log_{10} (T/T_K)$ calculated
by various methods.
The dots and thin solid line are for
NRG~\cite{Costi94} and Bethe ansatz~\cite{Konik02}, respectively.
The data is taken from Ref.~\cite{Konik02}.
The thick solid line is for the strong coupling perturbation theory.
The broken line for the weak coupling perturbation theory~\cite{Kaminski99}. }
  \label{fig:conductance}
\end{figure}
In Fig. 1,
the conductance  is plotted as a function of $\log_{10} (T/T_K)$,
calculated by various methods:
the weak coupling perturbation theory~\cite{Kaminski99}, NRG~\cite{Costi94}, Bethe ansatz~\cite{Konik02}, and
the strong coupling perturbation theory.
The data of NRG and Bethe ansatz is taken from Ref.~\cite{Konik02}.
The perturbative  strong coupling theory (the thick solid line) qualitatively coincides
with the results of NRG and Bethe ansatz (the dots and thin solid line)
near $T \lesssim  T_{K}$.
As $T$ increases above $T_{K}$,
the deviation from the results becomes larger. 
It is in contrast to the result of the weak coupling  theory, 
where
$G$ deviates from the results
when $T < 10 T_{K}$.
This difference describes the character of the Kondo screening cloud.
When $T > 10 T_{K}$, it is less prominent,
and grows during $T_{K} < T < 10 T_{K}$.
When $T < T_{K}$, it is fully developed.

Although the perturbation theory shows 
the unitary limit of the conductance,
it does not reproduce
the Fermi liquid picture at low temperatures;
$\log(T)$ behavior still appears.
This point is in contrast to  the conventional notion of
the strong coupling fixed point.
The discrepancy comes from the limit of the perturbative RG approach.
In the weak coupling theory,
the perturbation with respect to $J$ goes wrong as $D$ decreases
because it starts from the doublet state rather than the singlet state.
The same is true for the perturbation theory in the strong coupling regime.
The Fermi liquid picture is captured by
taking into account the spin singlet correlation appropriately
from the beginning, for example applying
a mean field theory for $J$~\cite{Read83}.

Nevertheless,
the strong coupling perturbation theory provides
a simple  approach to
the crossover regime between the Fermi liquid  and the weak coupling theories,
where has been treated only by more sophisticated or numerical methods.
Furthermore, it elucidates that
there is no transition around $T=T_{K}$, as has been shown.

In conclusion,
we have investigated the duality in the Kondo model to
identify the strong coupling Kondo model
with the coupling constant of $1/J$
in a simple manner.
The conductance calculated by
the perturbation theory with
the poorman's scaling
qualitatively agrees with the exact results near $T = T_{K}$.

The author acknowledges Y.~ Avishai, and Y. ~Meir for fruitful comments.
This work is supported by
the JSPS  Postdoctoral Fellowship.



\begin{thebibliography}{99}

\bibitem{Kondo64}
J. Kondo, 
Prog. Theor. Phys. {\bf 32}, 37 (1964).


\bibitem{Hewson92}
For a review, see A. C. Hewson,
{\it  The Kondo Problem to Heavy Fermions}
(Cambridge Univ. Press, Cambridge, 1992).
See also, J. Phys. Soc. Jpn., {\bf 74}, No. 1, (2005).


\bibitem{Goldhaber98}
D. Goldhaber-Gordon {\it et al.},
Nature {\bf 391}, 156 (1998);
Phys. Rev. Lett. {\bf 81}, 5225 (1998).

\bibitem{Cronenwett99}
S.M. Cronewett, T.H. Oosterkamp, L.P. Kouwenhoven, 
Science {\bf 281}, 540 (1998).

\bibitem{Simmel99}
F. Simmel {\it et al.}, 
Phys. Rev. Lett. {\bf 83}, 804 (1999).

\bibitem{Wilfred00}
W.G. van der Wiel {\it et al.},
Science, {\bf 289}, 2105
(2000).

\bibitem{Sasaki00}
S. Sasaki {\it et al.},
Nature {\bf 405}, 764 (2000).

\bibitem{Ji00}
 Y. Ji {\it et al.},
Science {\bf 290},779 (2000);
Phys. Rev. Lett. {\bf 88}, 076601 (2002).

\bibitem{Jeong01}
H. Jeong, A. M. Chang, and M. R. Melloch,
Science {\bf 293}, 2221, (2001).

\bibitem{Craig04}
N. J. Craig {\it et al.},
Science {\bf 304}, 565, (2004).


\bibitem{Fateev81}
V.\ A.\ Fateev, and P.\ B.\ Wiegmann,
Phys.\ Lett.\ A {\bf 81}, 179 (1981).

\bibitem{Fendley98}
P.\ Fendley, and H.\ Saleur,
Phys. Rev. Lett. {\bf 81}, 2518 (1998);
Phys. Rev. B {\bf 60}, 11432 (1999);
Nucl. Phys. {\bf B}, 571 (2000);
P.\ Fendley,
Adv. Theor. Math. Phys. {\bf 2}, 987 (1998).

\bibitem{Lesage99}
F.\ Lesage, and H.\ Saleur,
Phys. Rev. Lett. {\bf 82}, 4540 (1999);
Nucl. Phys. {\bf B} 546, 585 (1999).

\bibitem{Schmid83}
A. Schmid,
Phys. Rev. Lett. {\bf 51}, 1506 (1983). 

\bibitem{Fisher85}
M. P. A. Fisher, and W. Zwerger,
Phys. Rev. B {\bf 32}, 6190 (1985).

\bibitem{Weiss99}
U. Weiss,
{\it Quantum Dissipative Systems} (2nd ed.),
Chap.~25,
(World Scientific, Singapore, 1999).




\bibitem{Nozieres75}
Ph.~Nozi\`eres, in {\it Proc. 14th Int. Conf. on Low
Temperature Physics}, eds. M.~Krusius and M.~Vuorio, Vol. 5, (North
Holland, Amsterdam 1975).

\bibitem{Sorensen95}
 E.S. S\o rensen and I. Affleck, 
Phys. Rev. {\bf B51}, 16115 (1995).

\bibitem{Barzykin96} 
V. Barzykin and I. Affleck, Phys. Rev. Lett. 
{\bf 76}, 4959 (1996); Phys. Rev. {\bf B57}, 432 (1998).

\bibitem{Ujsaghy00}
 O. \'Ujs\'aghy, J. Kroha, L. Szunyogh and A. Zawadowski, 
Phys. Rev. Lett. {\bf 85}, 2557 (2000).


\bibitem{Zvyagin97}
 A.A. Zvyagin and P. Schlottmann, Phys. Rev. {\bf B54}, 
15191 (1997).

\bibitem{Ferrari99}
 V. Ferrari, G. Chiappe, E.V. Anda and M.A. Davidovich, 
Phys. Rev. Lett. {\bf 82}, 5088 (1999).

\bibitem{Kang00} K. Kang and S.-C. Shin, Phys. Rev. Lett. {\bf 85}, 5619 (2000).

\bibitem{Affleck01}
 I. Affleck and P. Simon, Phys. Rev. Lett. {\bf 86}, 2854
  (2001); P. Simon and I. Affleck, Phys. Rev. B {\bf 64}, 085308 (2001).

\bibitem{Hu01}
H. Hu, G.-M. Zhang, and L. Yu, 
Phys. Rev. Lett. {\bf 86}, {5558} (2001).

\bibitem{Eckle01}
 H.-P. Eckle, H. Johannesson and C.A. Stafford,
 Phys. Rev. Lett. {\bf 87}, 016602 (2001).

\bibitem{Simon02}
P. Simon and I. Affleck, Phys. Rev. Lett. 89, 206602 (2002);
Phys. Rev. B {\bf 68}, 115304 (2003)
  
\bibitem{Cornaglia03}
P. S. Cornaglia and C. A. Balseiro, Phys. Rev. Lett. {\bf 90}, 216801 (2003)

\bibitem{Sorensen05}
E. S. S{\o}rensen and I. Affleck,
Phys. Rev. Lett. {\bf 94}, 086601 (2005). 


\bibitem{Nozieres74}
Ph.~Nozi\`eres, J. Low Temp. Phys. {\bf 17}, 31 (1974).

\bibitem{Yamada75}
K. Yamada, Prog. Theor. Phys. {\bf 53}, 970 (1975); {\bf 54}, 316 (1975);
K. Yosida and K. Yamada,  {\bf 53}, 1286 (1975). 

\bibitem{Costi94}
T. A. Costi, A. C. Hewson, and V. Zlati\'c,
J. Phys. : Condens, Matter {\bf 6}, 2519 (1994).

\bibitem{Konik02}
R. M. Konik, H. Saleur, and A. W. W. Ludwig,
Phys. Rev. Lett. {\bf 87}, 236801 (2001);
Phys. Rev. B {\bf 66}, 125304 (2002).


\bibitem{Anderson70}
P. W. Anderson, J. Phys. C: Solid State Phys. {\bf 3}, 2436 (1970).


\bibitem{Glazman88}
L. I. Glazman and M. E. Raikh, JETP Lett. {\bf 47}, 452 (1988).

\bibitem{Kaminski99}
A. Kaminski, Yu. V. Nazarov, and L. I. Glazman,
Phys. Rev. Lett. {\bf 83}, 384 (1999);
Phys. Rev. B {\bf 62}, 8154 (2000).

\bibitem{Read83}
N. Read and D. M. Newns, 
J. Phys. C: Solid State Phys. {\bf 16}, 3273 (1983).

\end{thebibliography}
\end{document}